\documentclass{ws-ijmpa}

\begin{document}

\markboth{Korosteleva}{TUNKA}

\catchline{}{}{}{}{}

\title{Primary Energy Spectrum and Mass Composition Determined with the
 Tunka EAS Cherenkov Array}

\author{D.Chernov, E.Korosteleva, L.Kuzmichev, V.Prosin, I.Yashin}
\address{Scobeltsyn Institute of Nuclear Physics of MSU (Moscow, Russia)}

\author{N.Budnev, O.Gress, T.Gress, L.Pankov, Yu.Parfenov, Yu.Semeney}
\address{Institute of Applied Physics of ISU (Irkutsk, Russia)}  

\author{B.Lubsandorzhiev, P.Pohil}
\address{Institute of Nuclear Research of RAS (Moscow, Russia)}  

\author{T.Schmidt, C.Spiering, R.Wischnewski}
\address{DESY-Zeuthen (Zeuthen,Germany)}  

\maketitle


\begin{abstract}
 New results of 300 hours of operation of the Tunka array are presented. 
 An improved parametrization of the Cherenkov light lateral distribution  
 function (LDF), based on CORSIKA Monte Carlo simulations and the
 experiment QUEST, has
 been used for the reconstruction of EAS parameters. The corrected energy
 spectrum in the knee region is obtained. The mean depth of the EAS maximum has
 been 
 derived both from the analysis of LDF steepness and the FWHM of Cerenkov light
 pulse. The mean mass composition around the knee is estimated.
\end{abstract}

\keywords{energy spectrum; mass composition; EAS Cherenkov light.}

 
\section*{Experimental Differential Energy Spectrum.}
 The TUNKA EAS Cerenkov array is located in Tunka Valley, at an altitude of 675 m
 a.s.l., and was described in [1].
 The new fitting function (LDF) for the EAS Cerenkov light lateral distribution,
 derived from CORSIKA code simulation [2], has been applied
 to TUNKA data. 
 The primary energy $E_0$[$TeV$] has been obtained from the measured Cherenkov
 light 
 flux at a distance 175 m from the shower core $Q_{175}$[$photon\cdot cm^
 {-2}\cdot eV^{-1}$] with CORSIKA sumulated relation: $E_0 = 370\cdot
 Q_{175}^{0.96}$ 
The absolute energy calibration is based on the results obtained with the
QUEST experiment [3]. A Monte Carlo simulation of the experiment has shown that
the energy resolution is better than 18\%. 
   
\begin{figure}
\centerline{\psfig{file=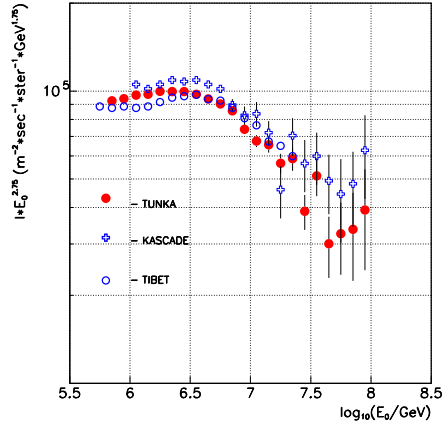,width=0.50\textwidth}}
  \caption{\textit{Differential energy spectrum.}}
\end{figure}

The spectrum is derived from data taken in 300 hours, spread over 51 clear
moonless nights, with a trigger rate of about 1.8 Hz.
To construct a spectrum with energy threshold $10^{15}$ eV, showers
with zenith angles $\theta \leq
25^{\circ}$ and a core position inside the geometrical area of the 
array have been selected.  
For the range from $6\cdot 10^{14}$ to $10^{15}$ eV,
events with zenith angles $\theta \leq 12^{\circ}$ falling inside a 5 times
smaller area around the array center have been selected. 
  
\section*{Estimation of Primary Mass Composition}

Lateral and time distributions of EAS Cherenkov light provide two independent
methods to estimate the maximum depth.
The simulation shows, that the LDF steepness $P = Q(100)/Q(200)$ is 
related to the linear distance (in [km]) from
the array to the EAS maximum position: $H_{max} = 10.62-0.12\cdot (P+2.73)^2$.
   
The Cherenkov light pulse FWHM [ns] at
distances larger than 200 m from the EAS axis is related to the relative position
of the EAS maximum by $\Delta X = X_0/cos\theta -X_{max}$ [g/cm$^2$], where $X_0$
is the total depth of the atmosphere and $\theta $ is the zenith angle of the
shower. This relation depends only on the distance to the EAS axis. 
For example, for distance of 250 m: $\Delta X = 1677+1006\cdot log_{10}(FWHM)$.
This method gives a better theoretical accuracy, than the first one.
Moreover, the $X_{max}$  estimate does not depend on assumptions 
about the primary nucleus. 

Figure 2 presents the mean depth of the EAS maximum, derived with the two methods
described above, as a function of primary energy. It is seen that the threshold
of the FWHM method is higher than that of the LDF steepness
method, but the mean depths, obtained with the two different methods are in good
agreement.

\begin{figure}[t]
\psfig{file=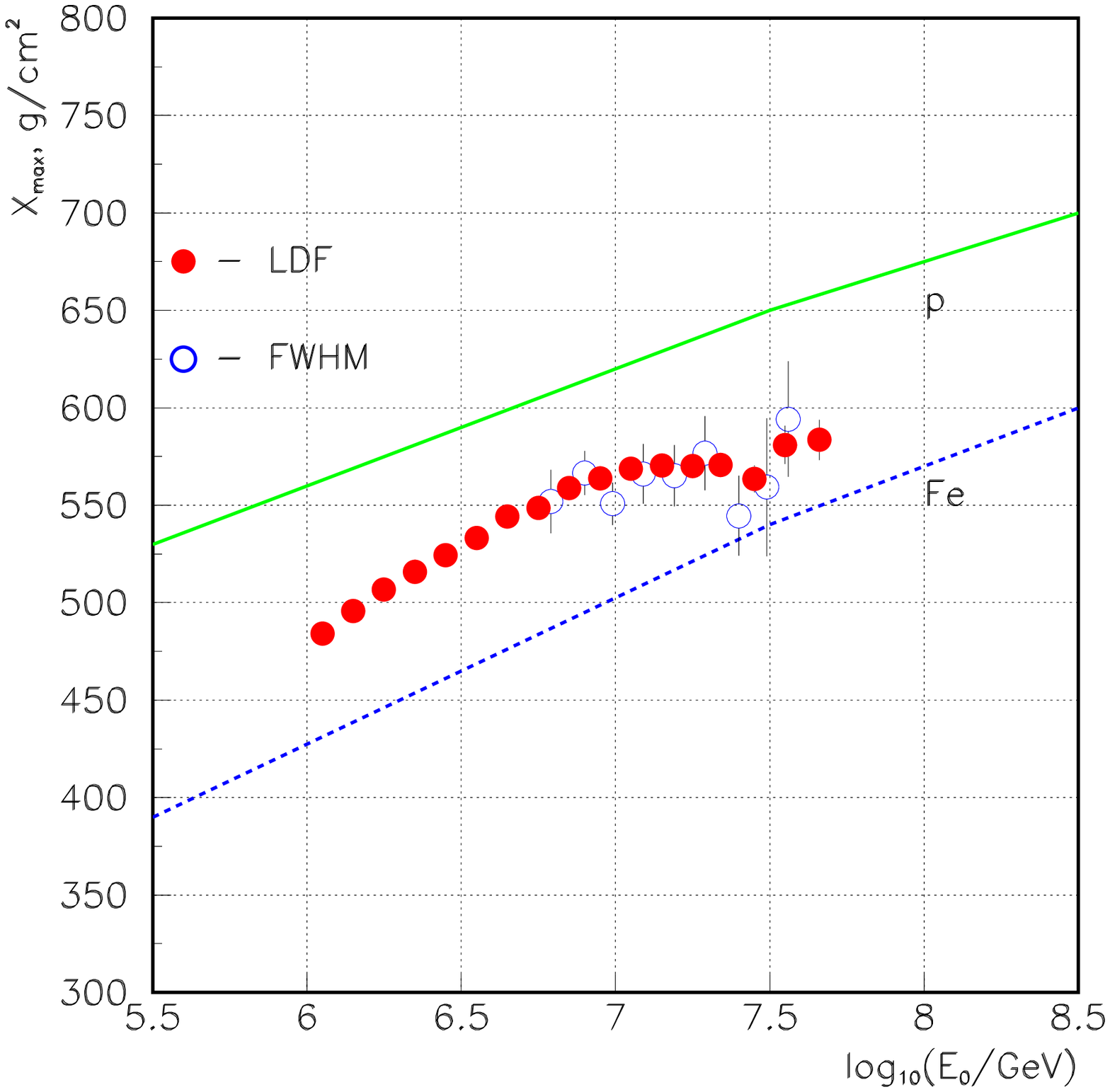,width=0.48\textwidth}
\hfill
\psfig{file=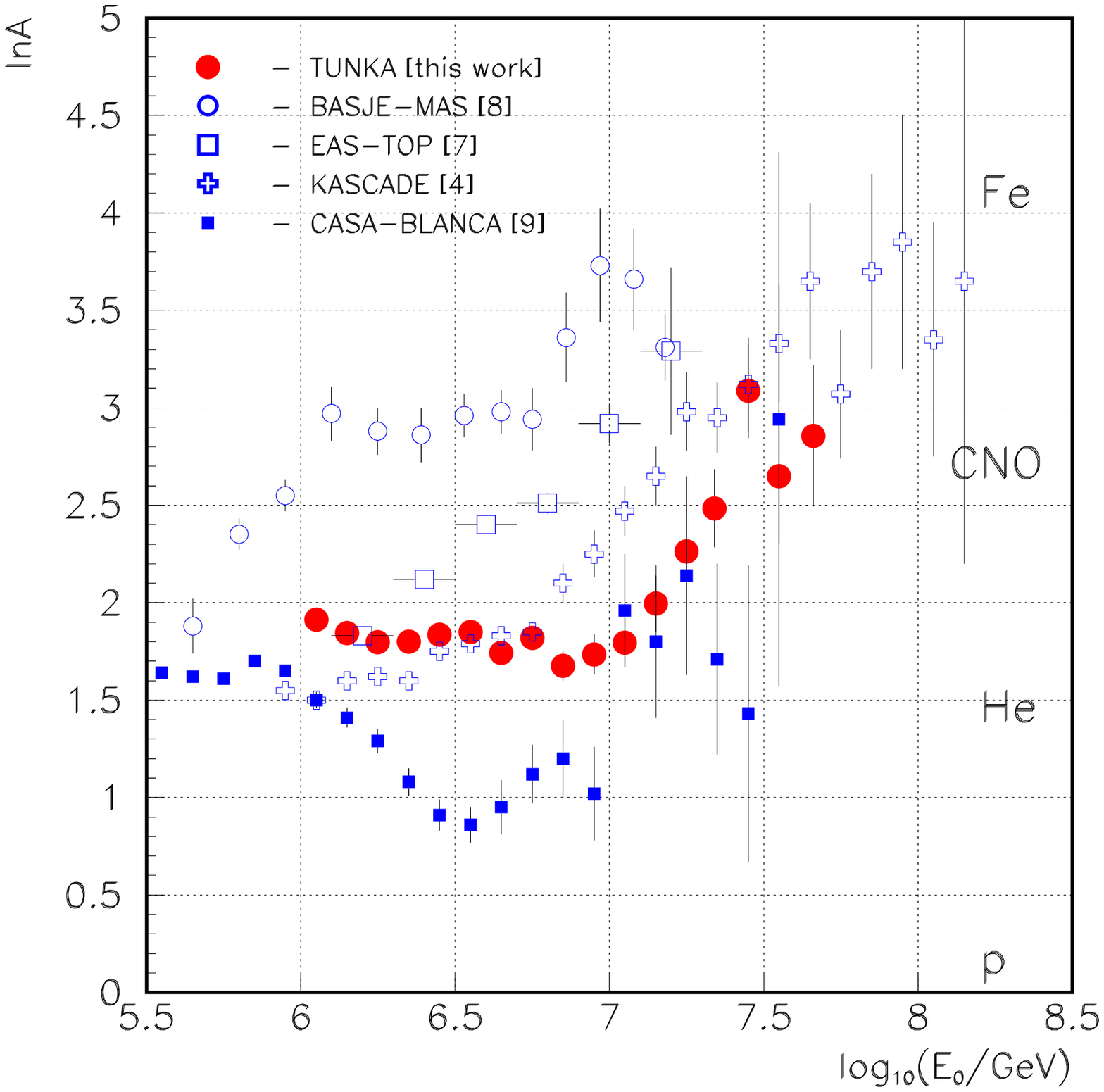,width=0.48\textwidth}
\vspace*{8pt}
\parbox[t]{0.48\textwidth}{\caption{Mean $X_{max}$
vs primary energy $E_0$. Curves are CORSIKA/QGSJET simulations.}}
\hfill
\parbox[t]{0.48\textwidth}{\caption{Mean $<lnA>$ vs primary energy.}}
\end{figure}

The mean values of $X_{max}$ from fig.2 can be easily
transformed to the mean logarithmic mass $<lnA>$ of primary particles. 
Figure 3 represents the result.
A slight correction, derived from MC simulations of the experiment (assuming a
4-group mass composition, p:He:CNO:Fe = 0.3:0.3:0.2:0.2) has been applied. 
According to these data the mass composition has almost no 
energy dependence in the range from $10^{15}$ to $10^{16}$ eV and 
is compatible with the hypothetical composition with
$<lnA> = 1.75$ mentioned above. 
At energies above $10^{16}$ eV, however, a steep increase of the
average mass is observed. A similar increase
in average mass is observed in almost all experiments.
Points of the present work coincide with KASCADE data [4] at 
the  energy of the knee.

\section*{Acknowledgements}
      
 Authors are thankful to professor Gianni Navarra and EAS-TOP Collaboration for 
 the opportunity to carry out the calibration experiment QUEST at the EAS-TOP
 array.  
 
 Work is supported  by 
 the Russian Fund of Basic Researchs (grant 02-02-17162).

\end{document}